\documentclass
[pre,twocolumn,showpacs,showkeys,nobibnotes,floatfix,superscriptaddress]
{revtex4}

\usepackage{amsmath,amssymb,array}
\usepackage{graphicx}

\newcommand\Rey{\ensuremath{\mathrm{Re}}}

\begin{document}

\title{Convective instability in inhomogeneous media: impulse response
  in the subcritical cylinder wake}

\author{C. Marais}
\affiliation{PMMH, UMR7636 CNRS, ESPCI ParisTech, UPMC,
                   University Denis Diderot, France}
\author{R. Godoy-Diana}
\email{ramiro@pmmh.espci.fr}
\affiliation{PMMH, UMR7636 CNRS, ESPCI ParisTech, UPMC,
                   University Denis Diderot, France}
\author{D. Barkley}
\affiliation{Mathematics Institute, University of Warwick, Coventry CV4 7AL
  United Kingdom}
\author{J.E. Wesfreid}
\affiliation{PMMH, UMR7636 CNRS, ESPCI ParisTech, UPMC,
                   University Denis Diderot, France}

\date{\today}

\begin{abstract}
  We study experimentally the impulse response of a cylinder wake below the
  critical Reynolds number of the B\'enard-von K\'arm\'an instability.  In this
  subcritical regime, a localized inhomogeneous region of convective
  instability exists which causes initial perturbations to be transiently
  amplified. The aim of this work is to quantify the evolution resulting from
  this convective instability using two-dimensional particle image velocimetry
  in a hydrodynamic tunnel experiment. The velocity fields allow us to
  describe the evolution of wave packets in terms of two control parameters:
  the Reynolds number and the magnitude of the imposed perturbation. The
  temporal evolution of energy exhibits a transient algebraic growth at short
  times followed by an exponential decay.
\end{abstract}



\maketitle


\section{Introduction}
\label{intro}

Flow past a circular cylinder is a classic prototype for studying hydrodynamic
instabilities and bifurcations in separated flows
\cite{Provansal1987,Jackson1987,Noack1994,Pier2002,Chomaz2005,
Barkley:2006,GiannettiLuchini:2007}. Moreover,
the academic case of a two-dimensional cylinder wake can be used as a basic
model for many real situations, including the flow behind support cables or
around an airfoil.  When the Reynolds number $\Rey={U_0D}/{\nu}$ (where $U_0$
is the free-stream velocity, $D$ is the cylinder diameter, and $\nu$ is the
kinematic viscosity) reaches a particular critical value ($\Rey_c\approx47$
for an infinitely long cylinder, \cite{Provansal1987,Jackson1987}), a
sustained periodic shedding of opposite-signed vortices gives rise to the
well-known B\'enard-von K\'arm\'an vortex street.

\begin{figure}
\centering
  \includegraphics[width=1\linewidth]{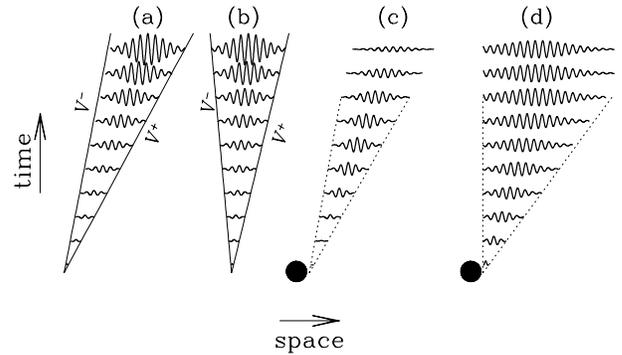}
  \caption{Sketch of the space-time response of flows to infinitesimal
    perturbations. (a) and (b) correspond to local analysis, i.e.\ parallel
    flows, illustrating the distinction between (a) convective and (b)
    absolute instabilities. The leading-edge velocity $V^+$ is positive in
    both cases. For convective instability, the trailing-edge velocity $V^-$
    is also positive, while for absolute instability it is negative.  (c) and
    (d) correspond to the global wake illustrating the distinction between the
    (c) subcritical, $\Rey < \Rey_c$, and (d) supercritical, $\Rey > \Rey_c$,
    cases. In the subcritical case, the perturbation reaches a maximum and
    subsequently decays. In the supercritical case the perturbation continues
    to grow until it saturates nonlinearly.  The edge velocities can be
    obtained over finite times only in experiment (see
    Sec.~\ref{sec:packet}).}
  \label{marais_intro}
\end{figure}

The transition to sustained oscillations can be described locally, via wake
profiles at different spatial stations, or globally, viewing the 2D wake as
whole. To elaborate further, we first recall the standard distinction between
convective and absolute instability in parallel flows illustrated in
Figs.~\ref{marais_intro}(a) and \ref{marais_intro}(b). An instability is
convective if a perturbation grows but is simultaneously advected with the
flow such that the disturbance decays at any fixed point, as in
Fig.~\ref{marais_intro}(a), while it is absolute if the perturbation grows at
a fixed spatial position, as in
Fig.~\ref{marais_intro}(b)~\cite{Landau1959,Huerre1985,hm90}.  A useful way to
distinguish convective from absolute instability is in terms of propagating
edge or front velocities: assuming a positive leading-edge velocity $V^+$, as
is the case here, convective instability corresponds to a positive
trailing-edge velocity $V^-$, while absolute instability corresponds to a
negative trailing-edge
velocity~\cite{Deissler1985,Huerre1985,hm90,VanSaarloos2003,Delbende1998}.

Local analysis of wake profiles, i.e.\ a parallel flow approximation, gives a
picture of the transition to sustained oscillations as
follows~\cite{Monkewitz1988,Delbende1998,Pier2002}.  Above $\Rey \simeq 5$
there is a local region of convectively unstable flow in the wake.  Above
$\Rey \simeq 25$ there is additionally a pocket of absolutely unstable
flow within the region of convective instability. Once the pocket of absolutely unstable flow becomes sufficiently large,
the flow becomes globally unstable.
Globally, however, one does not observe the onset of a locally absolutely
unstable region. Instead, one finds that below $\Rey_c$ the wake responds to
perturbations, but only transiently, as perturbations are advected through the
systems. This is illustrated in Fig.~\ref{marais_intro}(c).  Above $\Rey_c$,
as in Fig.~\ref{marais_intro}(d), perturbations grow and lead to a
synchronized wake in the formation of what is called a global
mode~\cite{Zielinska1995,Chomaz2005}.  Hence, even though convective and
absolute instability are strictly defined for
parallel flows and streamwise period flows~\cite{Huerre1985,hm90,Schatz1995},
inhomogeneous flows, such as flow past a cylinder,
may exhibit similar characteristics. The transient response in the subcritical
regime has the hallmarks of convective instability while the global
instability above $\Rey_c$ has the hallmarks of absolute instability.

Despite the large body of work on the cylinder wake, only a few experimental
studies have examined the wake's subcritical behavior~\cite{LeGal2000} and none
have {\em quantitatively} characterized the transient dynamics in this regime.
The goal of the present paper is therefore to analyze quantitatively the
subcritical regime using a well-controlled experiment.

After describing the experimental setup, we focus firstly on the evolution of
amplified wave packets and on obtaining the leading-edge, trailing-edge, and
group velocities in an experimental setting.  The decrease of trailing-edge
velocity towards zero when approaching the global instability threshold
confirms the transition to an absolute instability. The subcritical behavior
is further characterized in terms of the evolution of the maximum amplitude of
the wave packet, and its space-time position, as a function of the strength of
the perturbation and the distance to the B\'enard-von K\'arm\'an instability
threshold. Finally, to quantify the transient growth phenomenon due to the
inhomogeneity of the media, we analyze the temporal evolution of the energy of
the perturbation.

\begin{figure}
  \includegraphics[width=1\linewidth]{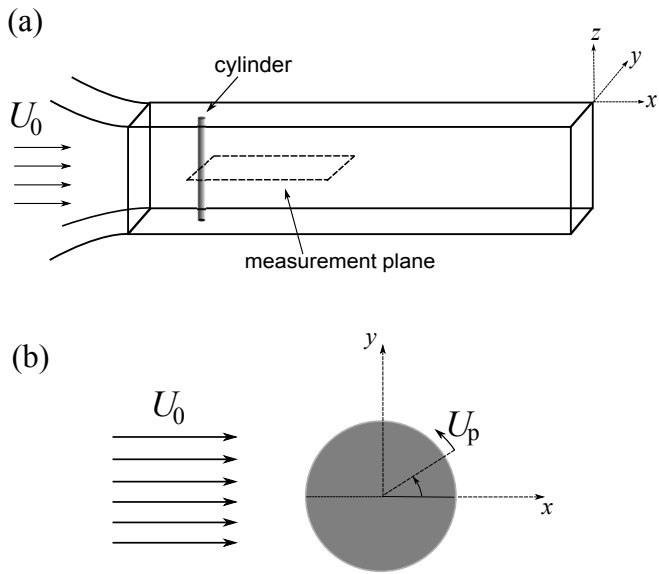}
  \caption{(a) Schematic diagram of the experimental setup in the hydrodynamic
    tunnel. (b) Sketch showing a top view of the cylinder during an impulsive
    rotation of intensity $U_{p}$. }
\label{marais_expe}
\end{figure}

\section{Experimental setup}
\label{sec:exp}

A cylinder of diameter $D = 5$ mm is placed in a hydrodynamic tunnel of
section 100mm x 100mm (see Fig.~\ref{marais_expe}) with a nearly plug flow
in the test section. The boundary layer width in the tunnel walls is of $\sim10$ mm
in the region of interest. The cylinder span is 98 mm
which covers practically the whole height of the tunnel.
We use a Cartesian coordinate system, placed in the cylinder center,
with the $x$-axis pointing in the flow direction and the $z$-axis running along
the cylinder centerline. The cylinder can be put into rotation to provide
impulse perturbations to the flow. Measurements
are taken in the horizontal mid-plane of the channel.

We define and work with the reduced Reynolds number
$\epsilon=(\Rey-\Rey_c)/\Rey_c$, corresponding to the distance from the global
instability threshold.  The critical Reynolds number measured in this
experiment is $\Rey_c\approx64$, which is larger than the ideal 2D case,
primarily due to confinement and blockage effects.  We adjust the Reynolds
number in the experiment by controlling the flow rate in the tunnel. In the
results that follow, $\epsilon$ varies from $-0.30$ to $-0.04$. For this range
of Reynolds numbers, the flow remains two-dimensional over nearly the entire
cylinder span.

Impulse perturbations consist of applying very short rotary motions to the
cylinder.  Rotation is controlled by a programmable microstepping electronic
module which gives a resolution of 1/256 per full step, allowing for a precise
control of the cylinder motion.  In the present work, we fix the
non-dimensional time interval over which the cylinder is rotated, and use the
speed of rotation as the amplitude of the perturbation.  More specifically,
the perturbation amplitude is defined from the tangential speed of rotation at
the cylinder surface $U_{p}$ (see Fig.~\ref{marais_expe}).  We consider three
perturbation amplitudes given by three values of the non-dimensional rotation
speed: $U_{p}/U_0=75, 100$ and $125$, which we refer
to as small, medium and strong, respectively. $U_0$ is the measured velocity
of the flat profile.  The small-amplitude perturbation is the smallest
perturbation that produced an observable response in the wake.  For all
applied perturbations, the cylinder is rotated for a fix dimensionless
duration given by $\triangle t/T_{adv}\approx0.2$, where $\triangle t$ is the
dimensional duration of the perturbation and $T_{adv}=D/U_0$.  Note that since
$U_0$ varies with Reynolds number, both the dimensional tangential speed
$U_{p}$ and the dimensional duration $\triangle t$, vary with Reynolds
number.



\begin{figure}
\centering
\includegraphics[width=\linewidth]{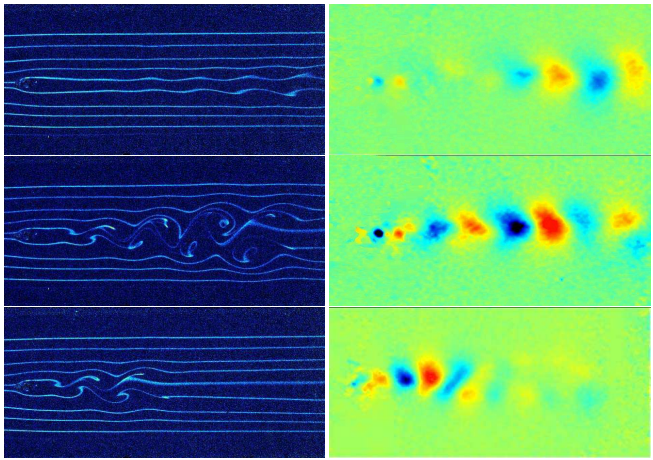}
  \caption{Visualization of the impulse response at three successive time
    instants (from bottom to top) in the cylinder wake.  Left: streaklines
    obtained from Fluorescein dye visualization. Right: instantaneous cross
    stream velocity field obtained from PIV measurements.  }
\label{marais_impulse}
\end{figure}

The wave packet generated by the perturbation can be observed qualitatively by
visualizing streaklines, as in the work of
\cite{LeGal2000}. Fig.~\ref{marais_impulse} shows this type of visualization
compared with snapshots of the cross-stream component of the velocity field
obtained by 2D Particle Image Velocimetry (PIV) in the horizontal mid-plane.
This highlights a significant difference between the present experiment and
previous studies of the subcritical wake response.  Velocity field
measurements obtained by PIV permit one to study directly the spatial
structure of velocity perturbations as they evolve.  The velocity
perturbations exhibit a well-defined maximum in the wake. These are
impossible to determine from streakline records since the deformation
amplitude of injected dye never decays downstream, an artifact caused by mass
conservation of the dyed fluid. Therefore, streakline deformation does not
give information about the amplitude of the velocity fluctuations $U_y$, and
thus about the spatial evolution of the impulse response.  PIV acquisition and
post-processing have been performed using a LaVision system with an ImagerPro
1600 x 1200 CCD camera with a 12-bit dynamic range capable of recording
double-frame pairs of images up to 15 Hz and a two rod Nd:YAG (15mJ) pulsed
laser. The time lapse between two frames is set to 20 ms. Finally, additional
post-processing and analysis has been carried out with Matlab and the PIVMat
Toolbox.

\section{Evolution of the wave packet}
\label{sec:packet}

\begin{figure*}
\centering
\includegraphics[width=0.48\linewidth]{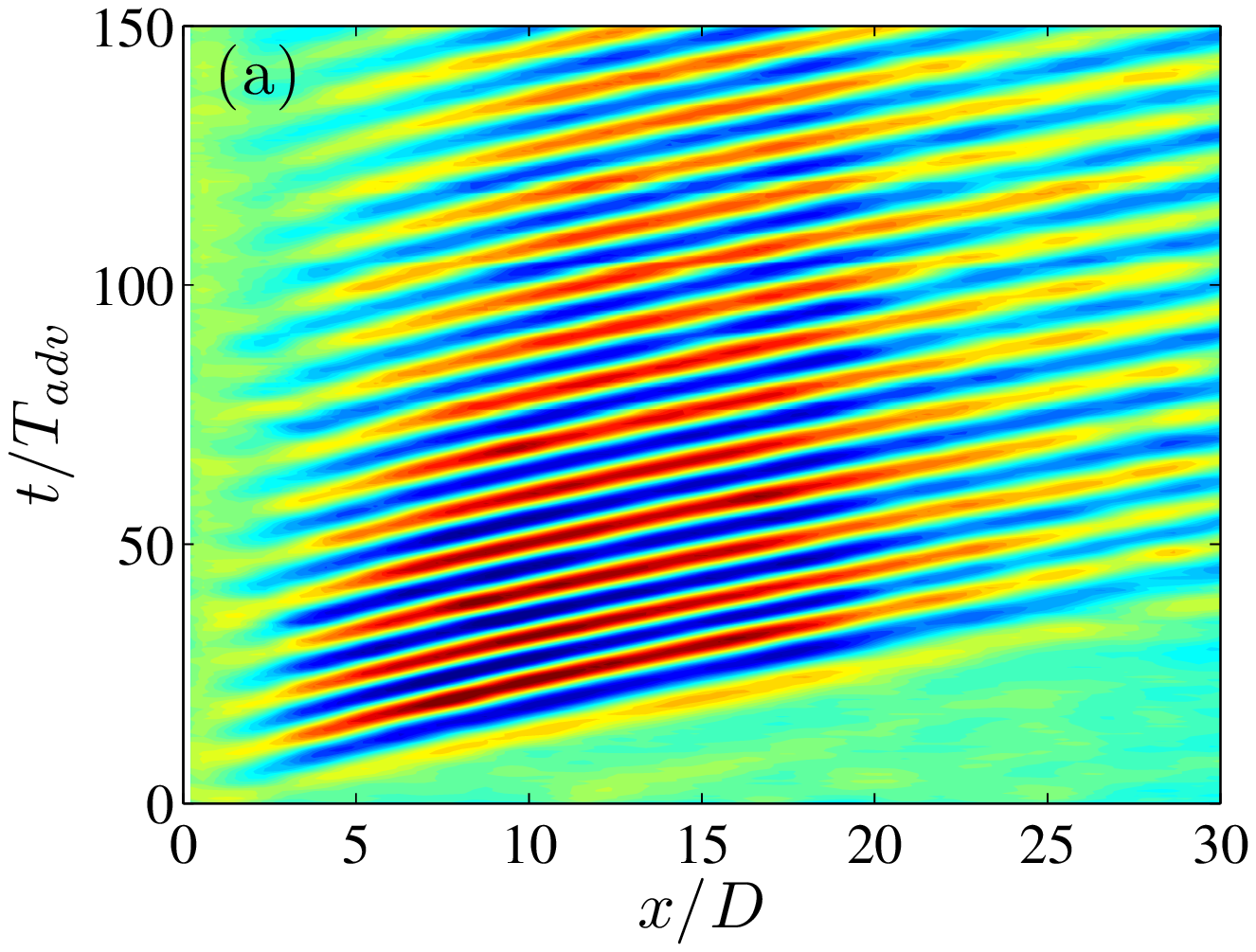}
\includegraphics[width=0.48\linewidth]{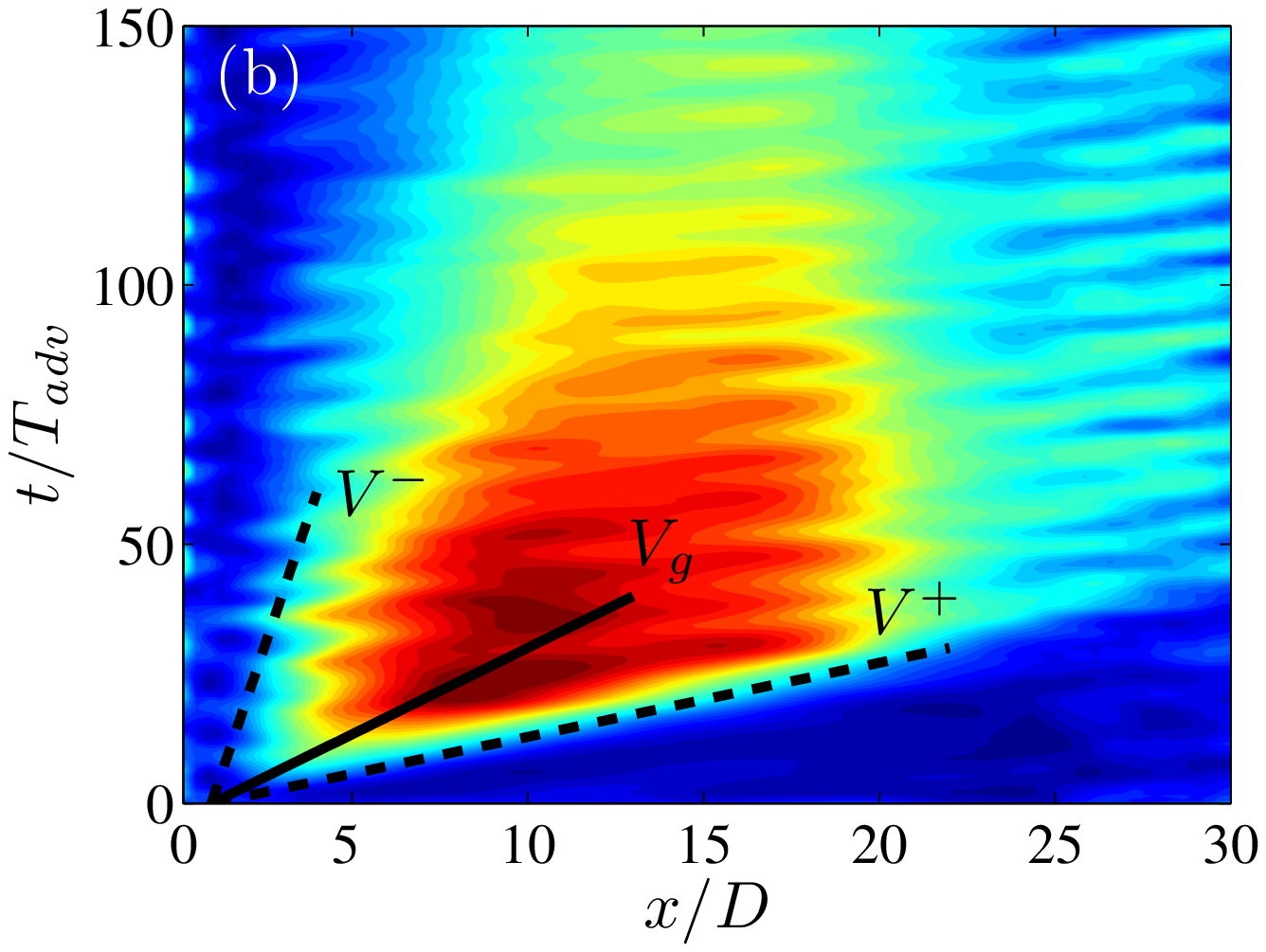}
\includegraphics[width=0.48\linewidth]{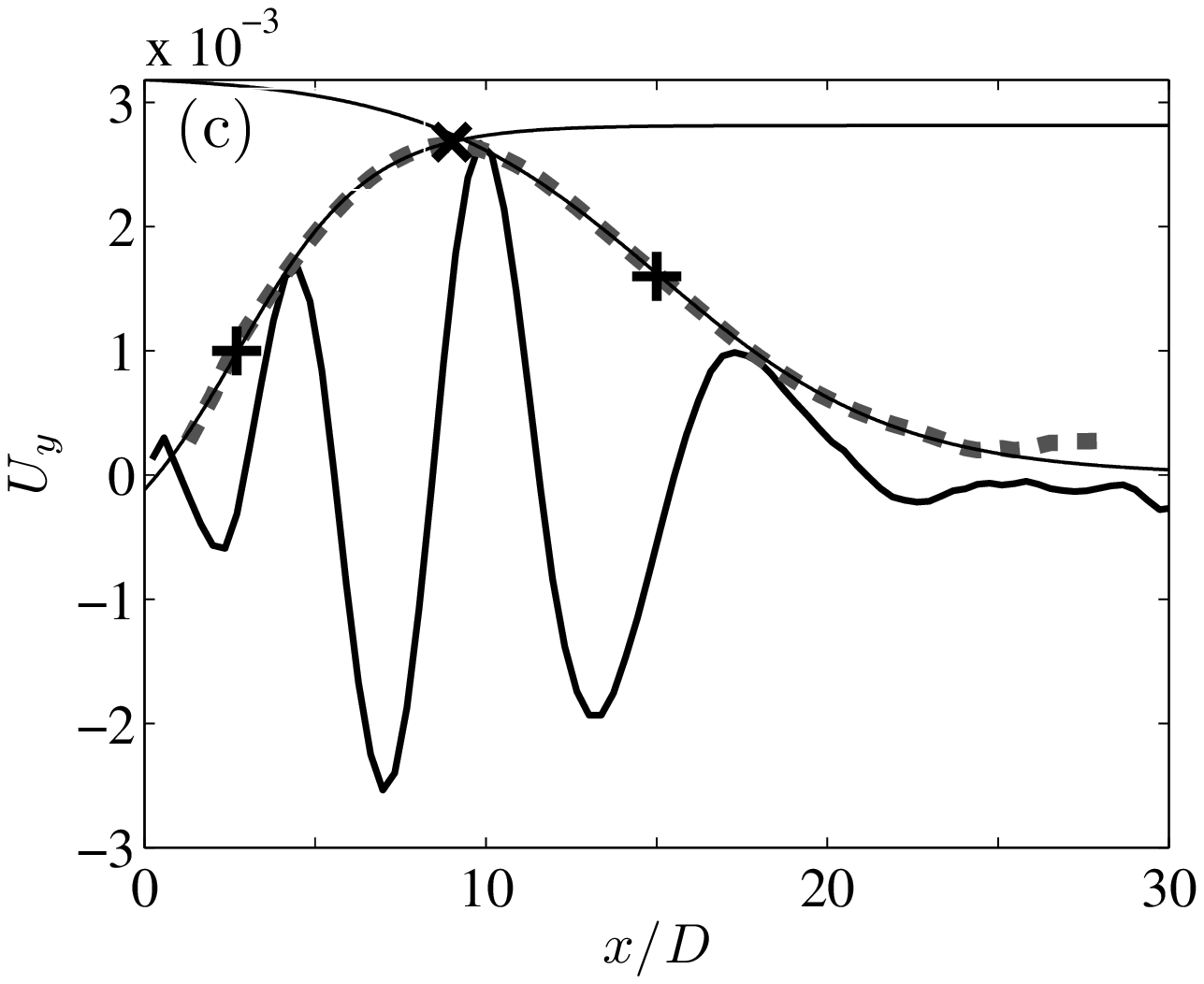}
\includegraphics[width=0.48\linewidth]{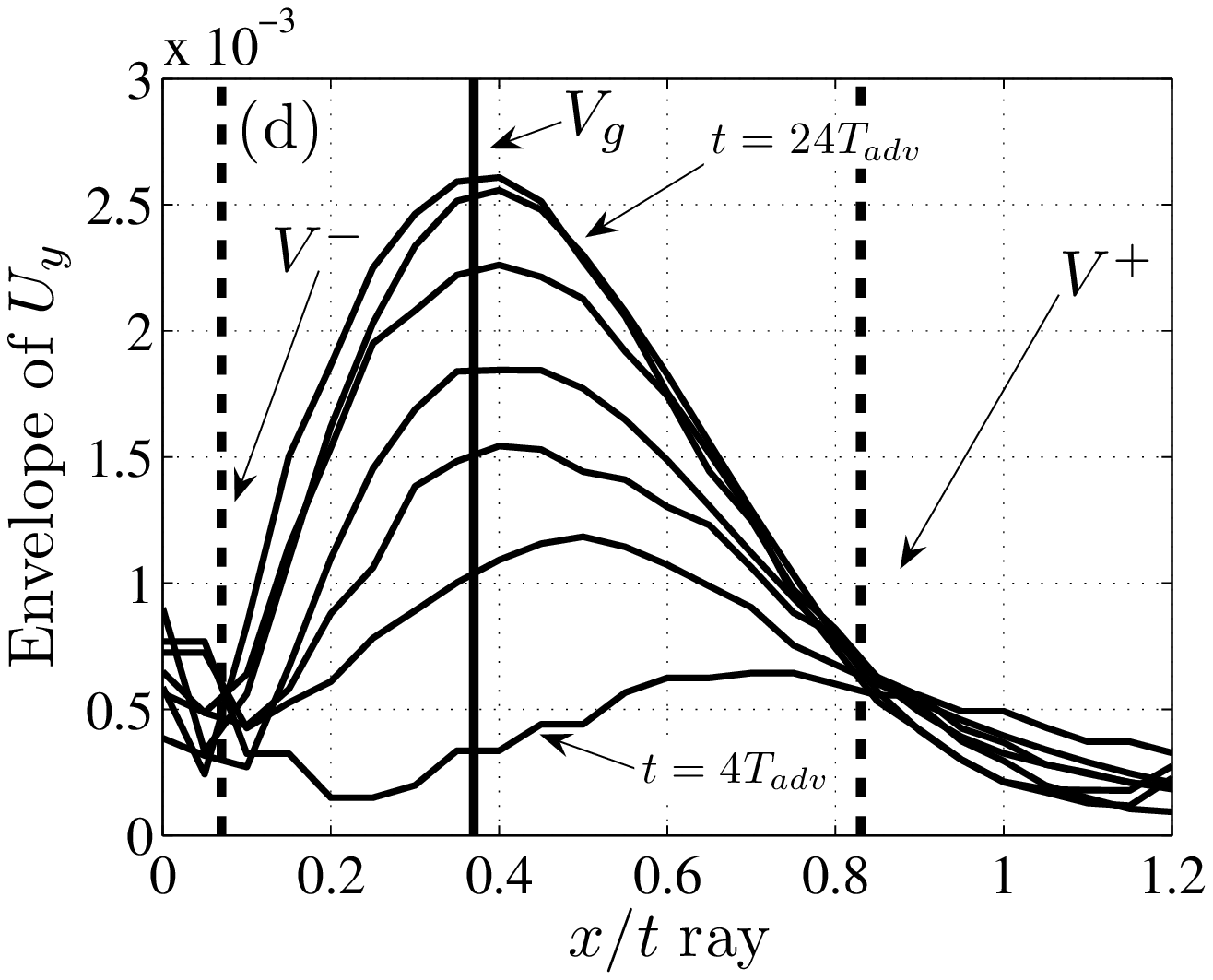}
  \caption{Evolution of the wave packet seen in spatiotemporal diagrams for
    $\epsilon=-0.137$. (a) Cross-stream velocity $U_y(x,t)$, (b) envelope of
    $U_y(x,t)$, (c) $U_y(x)$ and its envelope (bold dashed curve) at a fixed
    time ($t=24T_{adv}$). Also shown (thin curves) are $\tanh$ fits to the relevant part of
    the envelope. The infection points and envelope maximum are indicated.
    (d) Envelope of $U_y$ sampled along various rays as a function
    of ray velocity. The edge and group velocities, obtained from the
    inflection points and envelope maxima, are indicated with vertical
    lines.}
    \label{marais_spatio_velocities}
\end{figure*}

The convective nature of the impulsively perturbed cylinder wake is clearly
illustrated in the experimental data shown in
Figs.~\ref{marais_spatio_velocities}(a) and \ref{marais_spatio_velocities}(b).
In Fig.~\ref{marais_spatio_velocities}(a), cross-stream velocity $U_y$
profiles, measured at successive times along the wake symmetry axis $y=0$, are
stacked up to form a spatio-temporal diagram.  The cross-stream velocity on
the symmetry axis is the ideal quantity to use to investigate the perturbed
flow field because it is everywhere zero for the unperturbed flow.
Figure~\ref{marais_spatio_velocities}(b) is similar to
Fig.~\ref{marais_spatio_velocities}(a) except that the envelope [obtained
through a Hilbert transform of $U_y$ as illustrated in
Fig.~\ref{marais_spatio_velocities}(c)] is plotted.

We measure a number of velocities associated with the space-time
evolution of the wave packets.  From the Hilbert transform of the cross-stream
velocity we are able to extract, at each time instant, the envelope of the wave
packet as illustrated in Fig.~\ref{marais_spatio_velocities}(c).
Procedurally, we define the leading and trailing positions as the inflection
points of hyperbolic tangent fits to the relevant parts of the envelope at
each time. Such fits, together with the associated inflection points, are
included in Fig.~\ref{marais_spatio_velocities}(c).
Then we obtain the leading-edge velocity $V^{+}$ and the
trailing-edge velocity $V^{-}$ from the collection of fronts extracted from
the envelopes at short times ($\lesssim 50T_{adv}$).
Likewise, we obtain the envelope maximum at each time and from these data we
define the group velocity $V_g$ to be the velocity of envelope maximum, as
this gives the speed of the packet as a whole.
These velocities are the inverse of the slopes shown in the spatiotemporal
diagram of Fig.~\ref{marais_spatio_velocities}(b).
Finally, we define the phase velocity $V_p$ as the translation speed of the
vortices in the evolving wave packet as seen in
Fig.~\ref{marais_spatio_velocities}(a). Note that all vortices move at
essentially identical speed.
Moreover, the leading-edge velocity $V^{+}$, which is effectively determined
by the velocity of the first advected vortex released from the cylinder, is
the same as the phase velocity to within experimental precision: $V^{+} \simeq
V_p$.

We show now that the edge velocities obtained procedurally through inflections
points of the envelop correspond very closely to the edge velocities defined
in convectively unstable systems. (See in particular
Ref.~\onlinecite{Delbende1998}.) In this case one defines $V^-$ and $V^+$ to
be the $x/t$ rays that separate regions of growth from regions of decay in the
space-time diagram. The growth rate along rays defining the edges is zero.
For the experimental data, the growth rate along rays is shown in
Fig.~\ref{marais_spatio_velocities}(d), where the envelope of $U_y$, sampled
along different $x/t$ rays, is plotted against ray velocity.  The edge
velocities $V^+$ and $V^-$ obtained from inflection points are indicated by
vertical dashed lines in Fig.~\ref{marais_spatio_velocities}(d).  It can be
seen that these are in excellent agreement with the zero growth-rate
rays. Moreover, the group velocity $V_g$ (vertical solid line) is also agrees
very well with the maximum of the envelop in the ray representation.

Thus it can be seen that the velocities are all quite well defined
experimentally, as least up to the time at which the perturbation reaches its
maximum, even though the flow is spatially inhomogeneous and the wave packet
lives only a finite time.
As expected none of the velocities (including $V_p$) are strictly
constant over this space-time region, but they are very nearly so outside of
the near-wake region. We have not attempted to extract their detailed
variation in the present study.

\begin{figure}
\centering
\includegraphics[width=0.96\linewidth]{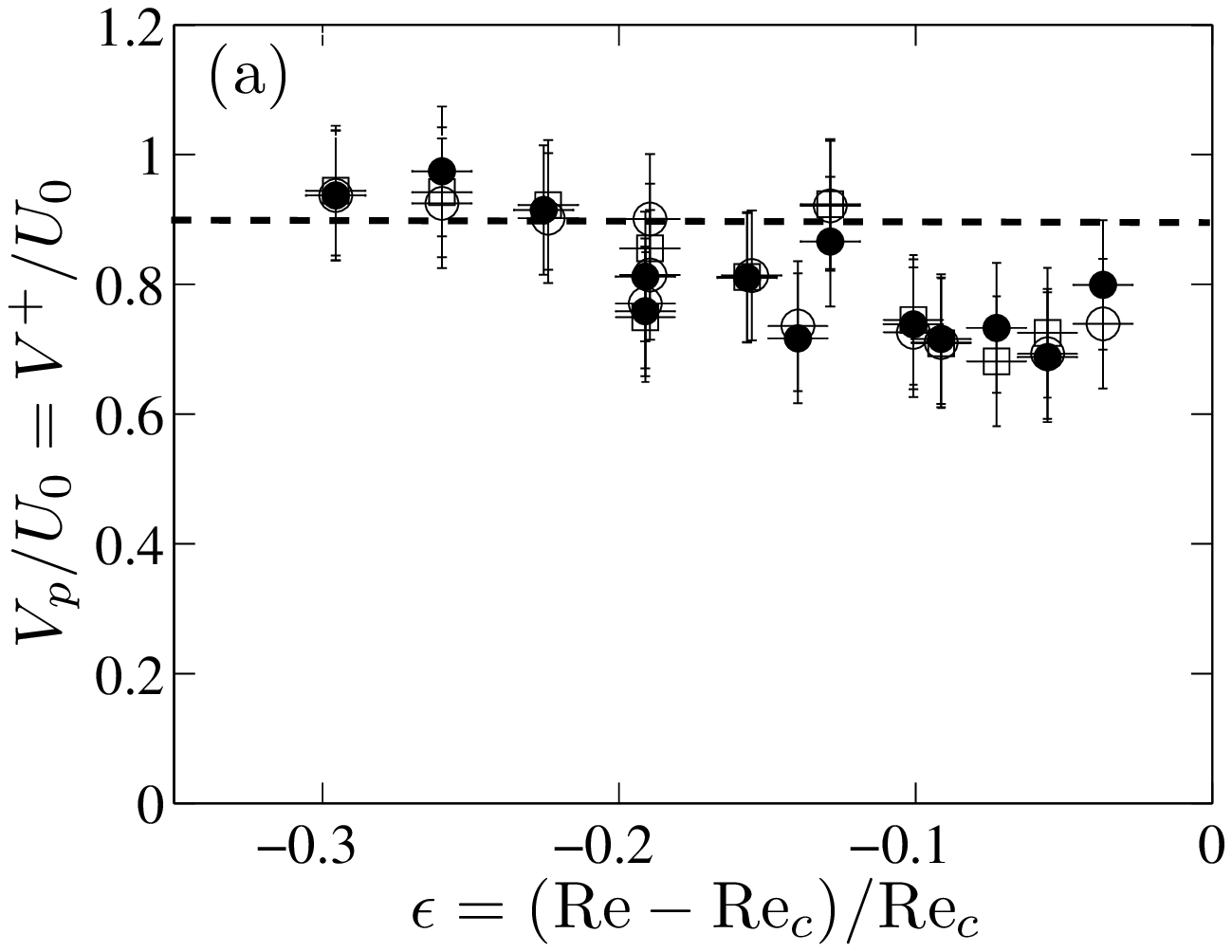}
\includegraphics[width=0.96\linewidth]{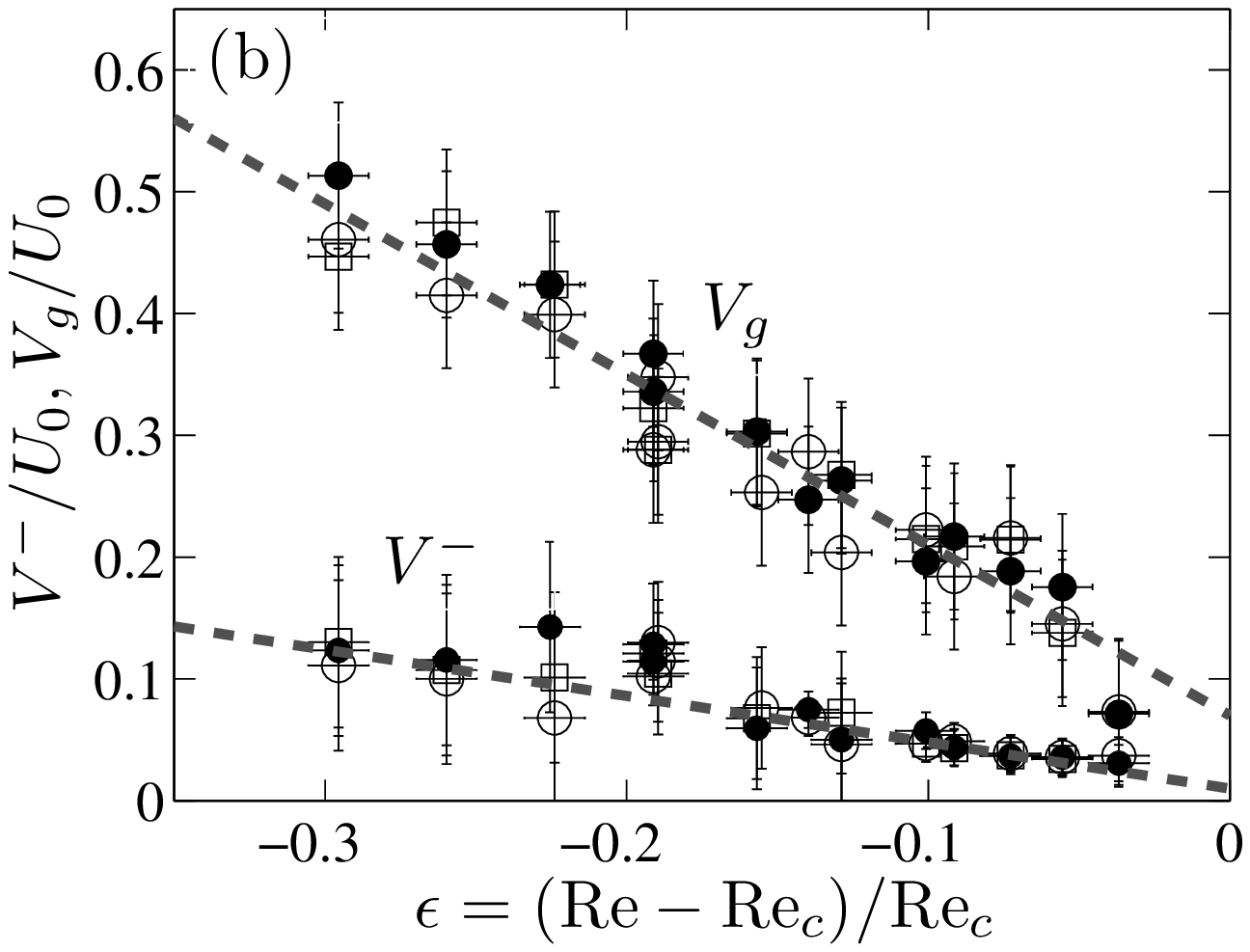}
  \caption{(a) Phase velocity $V_p$ (which is also $V^+$) and (b) group
    velocity $V_g$ and trailing-edge velocity $V^-$ as a function of
    $\epsilon$. The dashed line in (a) corresponds to the value of $V_p/U_0$
    for the B\'enard-von K\'arm\'an wake ($\Rey>\Rey_c$). The dashed lines in
    (b) are linear fits to the $V_g$ and $V^-$ data. Experimental points are
    labeled as $\bullet$: small perturbation, $\circ$: medium perturbation,
    $\square$: strong perturbation. Horizontal error bars indicate the
    uncertainty in the Reynolds number measurement.}
  \label{marais_spatio_velocities_vs_epsilon}
\end{figure}

We now consider the behavior of the different velocities as
$|\epsilon|\rightarrow0$ (i.e when $\Rey \rightarrow \Rey_c$).  In
Fig.~\ref{marais_spatio_velocities_vs_epsilon}(a) we can see that the phase
velocity $V_p$, and equivalently the leading edge velocity $V^+$, normalized
by the inflow speed, exhibit a slight decrease with increasing Reynolds
number.  Generally, the value of $V_p/U_0$ in the subcritical regime
$(\epsilon < 0)$ compares well with the value of $0.88$ reported in the
literature~\cite{Williamson1989} for the supercritical $(\epsilon > 0)$
regime.
The group and trailing edge velocities are summarized in
Fig.~\ref{marais_spatio_velocities_vs_epsilon}(b).  Both velocities decrease
with increasing Reynolds number as one would expect in approaching the global
wake instability.  To within experimental accuracy, $V^-$ approaches zero as
$\epsilon \to 0$.

\begin{figure*}
\centering
\includegraphics[width=0.48\linewidth]{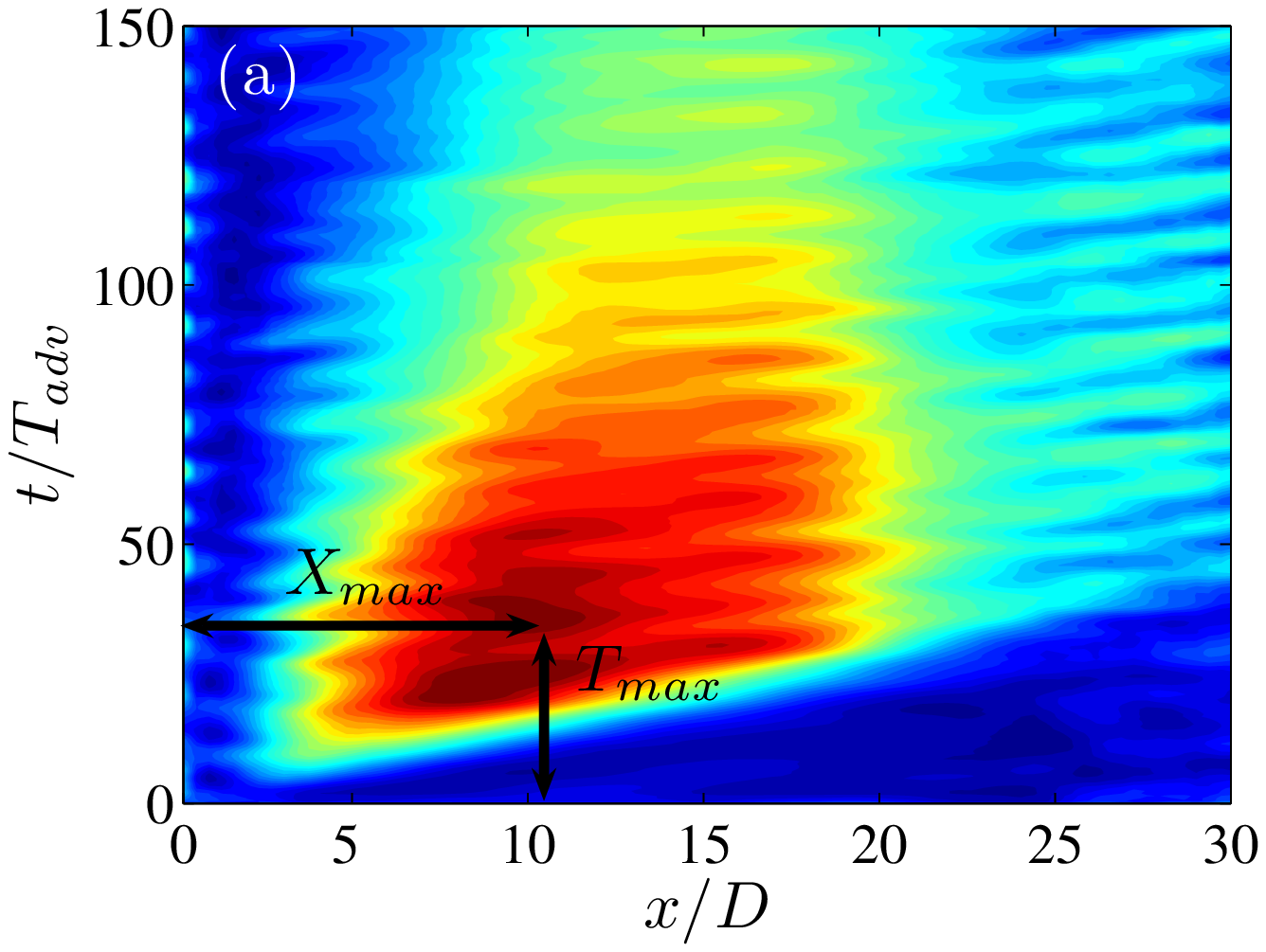}
\includegraphics[width=0.48\linewidth]{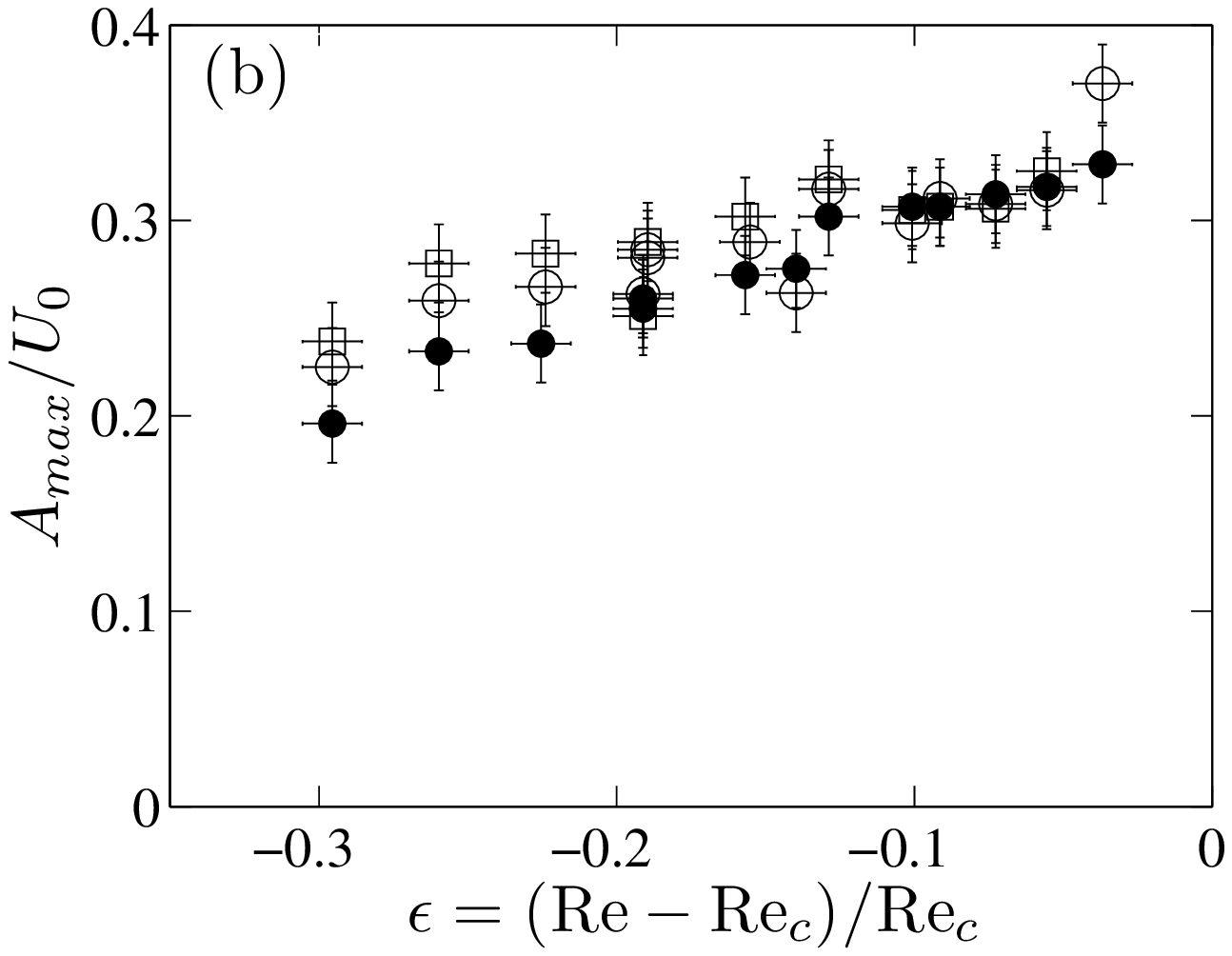}
\includegraphics[width=0.48\linewidth]{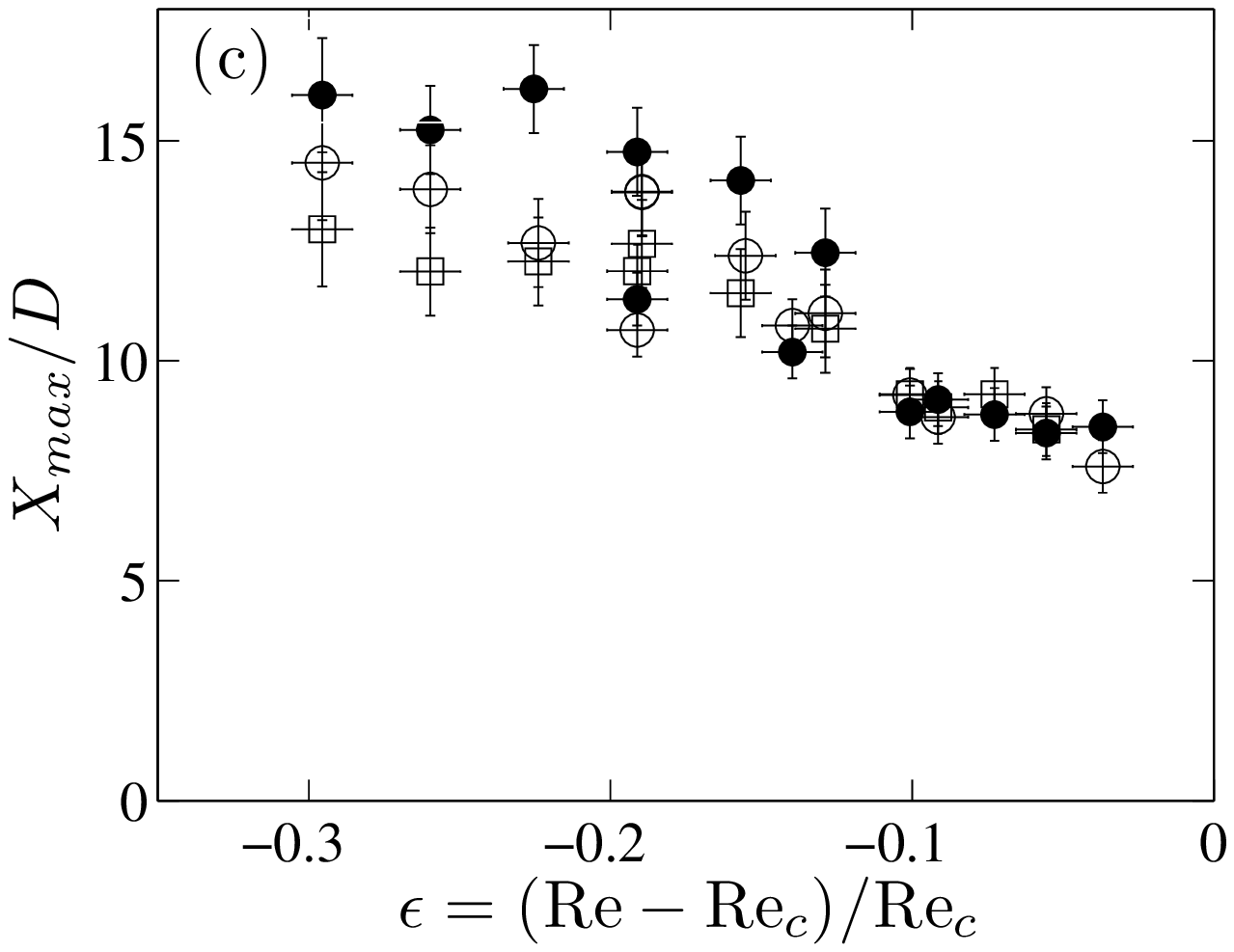}
\includegraphics[width=0.48\linewidth]{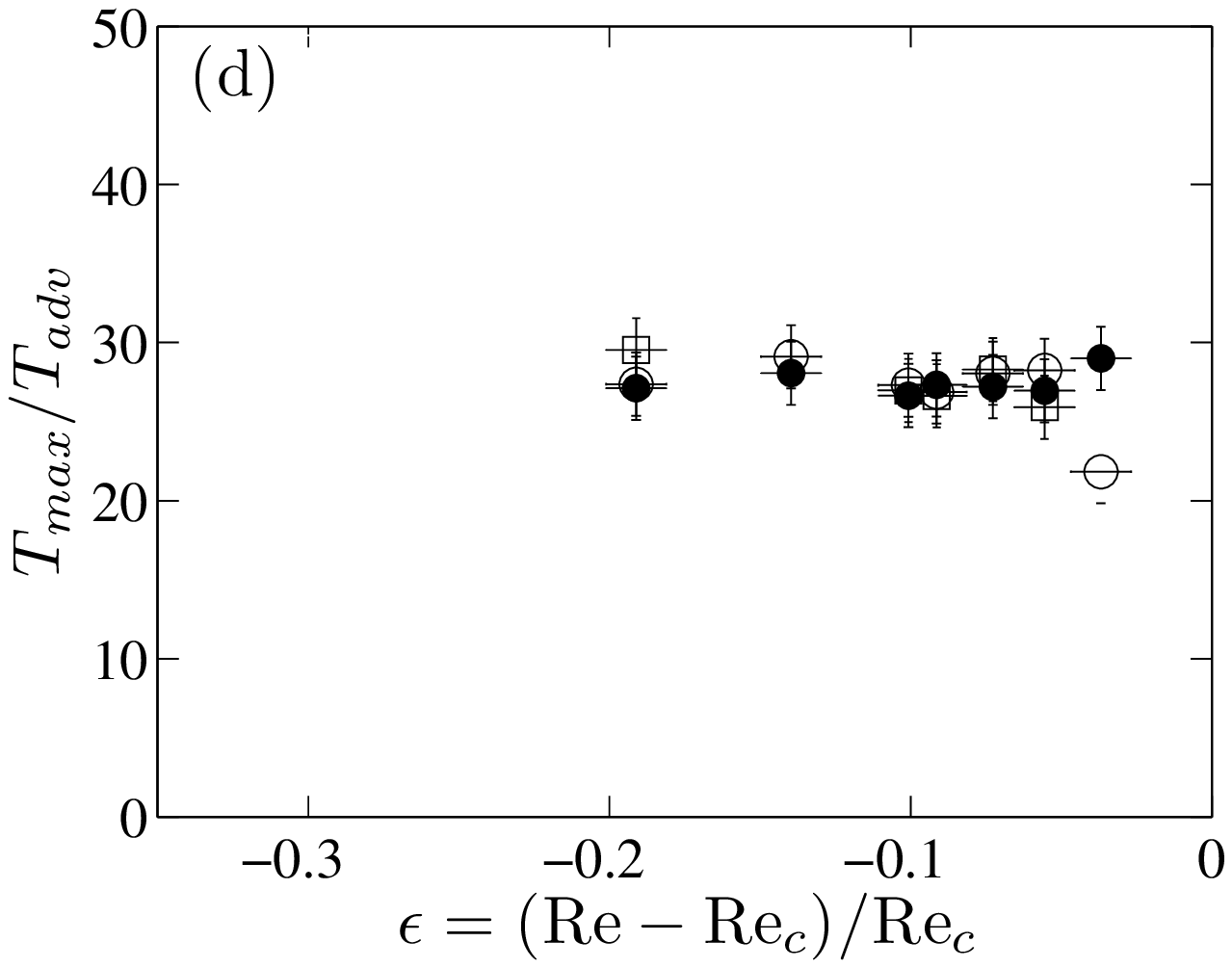}
  \caption{(a) Definition of the maximum of the perturbation on the
    spatiotemporal diagram. (b) Evolution of the maximum of the perturbation
    $A_{max}$, (c) $X_{max}$ and (d) $T_{max}$ as a function of the reduced
    Reynolds Number $\epsilon$. $A_{max}, X_{max}$ and $T_{max}$ are rendered
    nondimensionalized respectively by the free stream velocity $U_0$, the
    cylinder diameter $D$ and the advective timescale $T_{adv}=D/U_0$.
    Experimental points are labeled as $\bullet$: low perturbation, $\circ$:
    medium perturbation, $\square$: strong perturbation.}
    \label{marais_max}
\end{figure*}


From the spatiotemporal diagrams we are also able to pinpoint the position in
space and time $(X_{max},T_{max})$ at which the subcritical response reaches
maximum amplitude. See Fig.~\ref{marais_max}(a).  Again we use the
cross-stream velocity component $U_y$.  We let $A_{max}$ denote the maximum of
the response, so that $A_{max} = U_y(X_{max},T_{max})$.
Figures~\ref{marais_max}(b) through \ref{marais_max}(d) show the dependence of
the maximum on reduced Reynolds number.  It is not surprising that $A_{max}$
grows when the global instability threshold is approached
($|\epsilon|\rightarrow0$), as can be seen in Fig.~\ref{marais_max}(b),
because the susceptibility of the flow increases near the onset.  On the
contrary, the behavior of $X_{max}$ (which diminishes when
$|\epsilon|\rightarrow0$, see Fig.~\ref{marais_max}(c), deserves a
further comment, since it seems to be opposite to the case of other
instabilities in the subcritical regime where the characteristic length scale
increases when $|\epsilon|\rightarrow0$ (e.g. the penetration length in
pre-transitional Rayleigh-B\'enard convection \cite{Wesfreid1979}).  The fact
that the maximum of the instability moves closer to the cylinder (i.e. that
$X_{max}$ diminishes) when approaching the threshold while the time at which
this occurs remains constant [on a nondimensional scale normalized by the
advective time scale $T_{adv}$ , Fig.~\ref{marais_max}(d)] is actually
consistent with $V_g$ decreasing as the convectively unstable system tends to
the absolute instability threshold. The difference with the case of
Rayleigh-B\'enard convection comes from the effect of the mean flow advection,
which modifies the physical meaning of the penetration lengthscale.

For a given value of $\epsilon$, an increase in the strength of the
perturbation produces a response in which the maximum amplitude $A_{max}$ is
slightly larger and occurs at a position closer to the cylinder (smaller
$X_{max}$). Approaching the threshold reduces the effect of changing the
perturbation strength.

\section{Transient growth}

As noted in the introduction, an important feature of the cylinder wake is the
inhomogeneous nature of the flow.  Hence, even though the subcritical response
just presented has much of the general character of convective instability,
perturbations do not grow indefinitely (even linearly) as they would for a
homogeneous convectively unstable system.  Rather, a localized initial
perturbation grows at first, due to local flow features near the cylinder, but
is simultaneously advected downstream into a region of stability where the
perturbation decays.  Hence in the absence of any inflow noise the impulsive
response is only transient.  Such behavior is known in inhomogeneous
flows~\cite{cossu97,Gondret1999,Chomaz2005,Moehlis2006,Blackburn2008,
Marquet2008, Abdessemed2009, Cantwell2010a, Cantwell2010b}.
In should be noted, however, that inflow or other noise may
modify the picture in that the localized region of instability acts as an
amplifier and sustained dynamics may arise in some cases even in the
subcritical regime~\cite{Deissler1985, Gauthier1999, Marquet2008, Lopez2009,
Cantwell2010b}.

An increasingly common approach to quantifying the transient response of flows
is in terms of their transient energy growth~\cite{sh01}. Such an analysis
provides a global measure of the response dynamics. Here we undertake such an
analysis of experimental data.

We define the perturbation energy from our measurements as follows:
\begin{align*}
E(t) & = \int\int_{\Delta} \left( u_x^{2} + u_y^{2}\right)  dxdy
\end{align*}
where
\begin{align*}
u_x = U_x-U_{xbase} \quad u_y = U_y-U_{ybase} \; ,
\end{align*}
\noindent
where the base flow $(U_{xbase},U_{ybase})$ is the measured steady flow before
any perturbation.

\begin{figure*}
\centering
\includegraphics[width=0.32\linewidth]{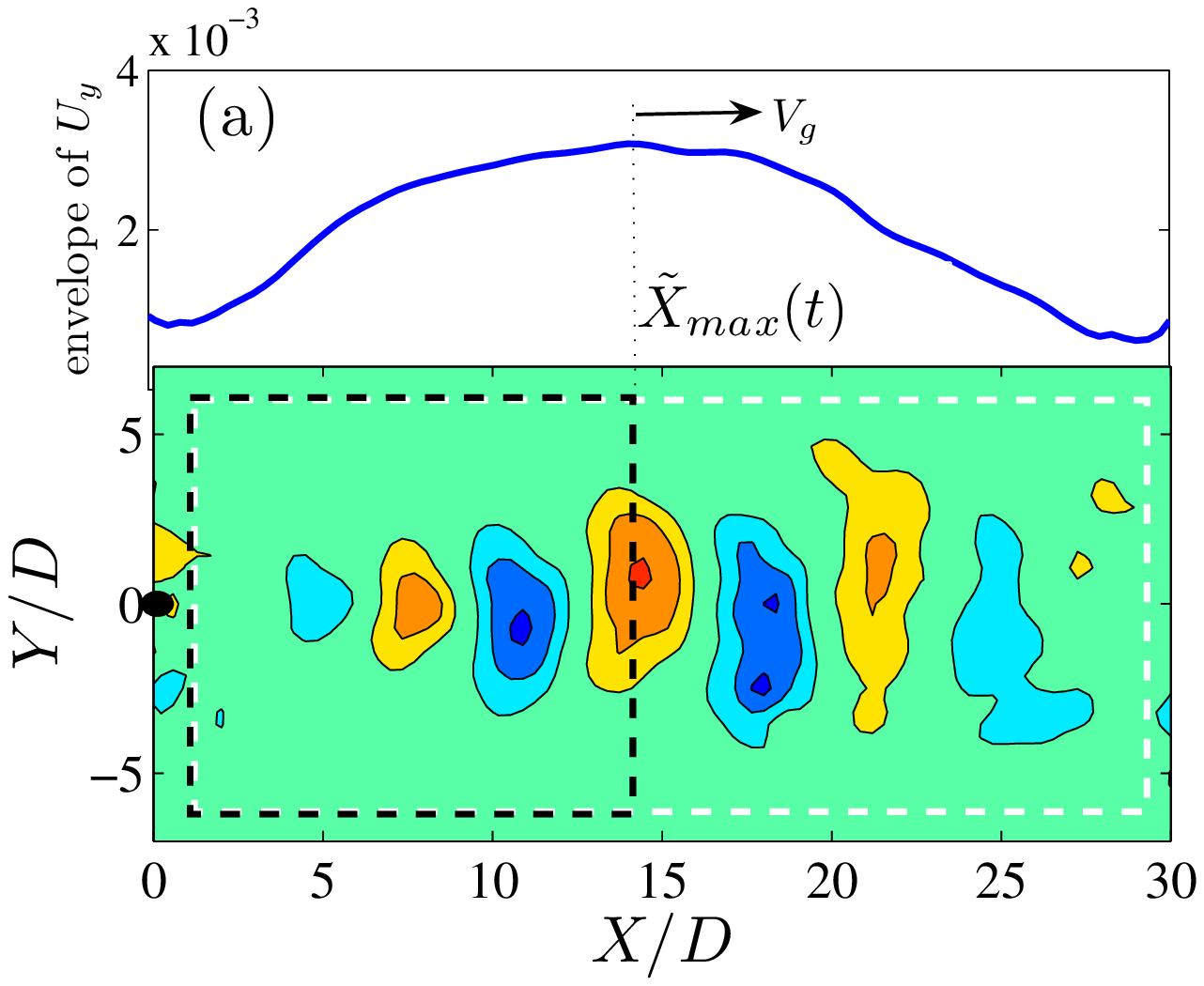}
\includegraphics[width=0.32\linewidth]{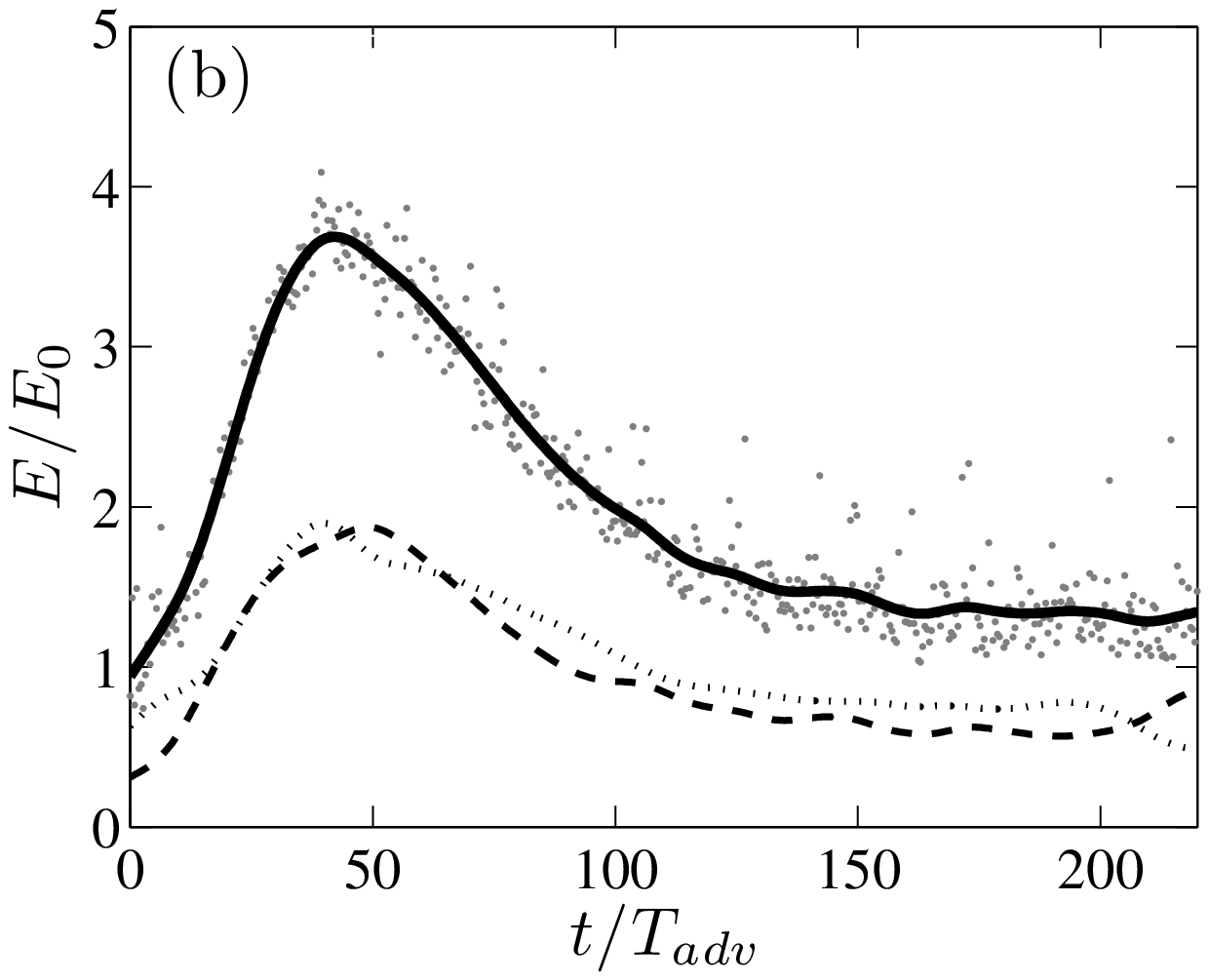}
\includegraphics[width=0.32\linewidth]{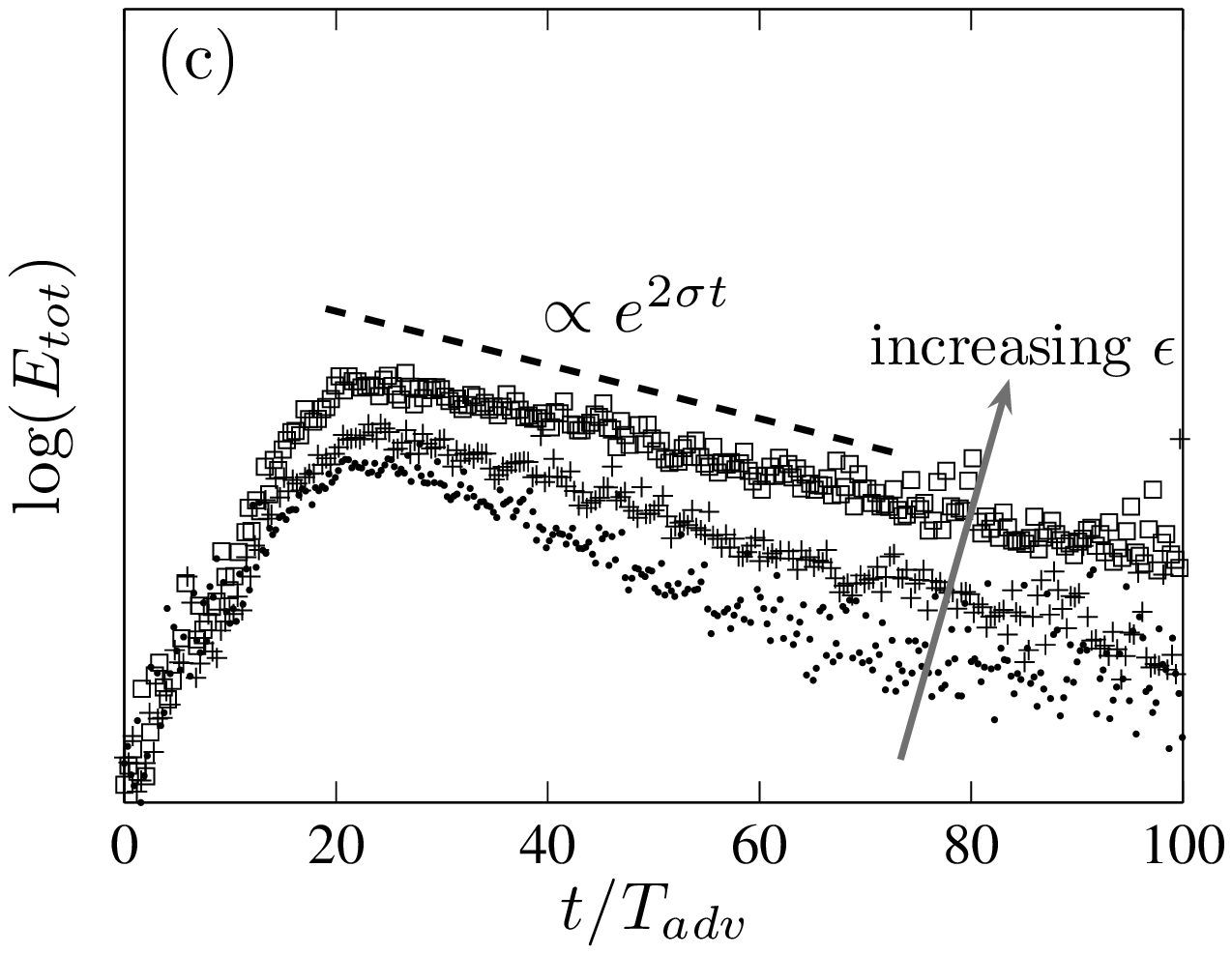}
  \caption{Transient energy growth in experiment. (a) Illustration of the
    integration area for the energy on the instantaneous
    velocity field $U_y$. $E_{tot}$ and $E_{half}$ are calculated over the
    white and black rectangles, respectively. The downstream boundary of the
    black rectangle moves at a speed $V_g$ and tracks the maximum of the
    perturbation envelope as indicated in the upper plot. (b) Energy as a
    function of time for $\epsilon=-0.14$.  The dots are the experimental data
    of $E_{tot}$. The solid, dashed and dotted lines are smooth fits of
    $E_{tot}$, $E_{half}$ and $E_{tot}-E_{half}$, respectively. (c) Time
    evolution of the energy on logarithmic scale, for three different values
    of the reduced Reynolds number $\epsilon$: $\bullet$ $\epsilon=-0.19$, $+$
    $\epsilon=-0.14$ and $\square$ $\epsilon=-0.1$.  The slope of the energy
    decay on the logarithmic scale gives the decay rate $\sigma$. }
  \label{marais_nrj}
\end{figure*}

Experimentally, the total energy of the perturbation can only be measured
while the packet is contained in the observation window. Hence, the energy
calculated from the velocimetry data does not include the contribution from
vortices that have been advected out of the measurement area.  In order to
quantify this effect we compare the energy calculated using two different
streamwise sizes for the integration area $\Delta$: the total energy
$E_{tot}$, where the whole measurement window is used, and the energy of the
first half of the wave packet $E_{half}$, where the downstream boundary of the
integration area $\Delta$ is given by the time-dependent position of the
maximum perturbation amplitude $\tilde{X}_{max}(t)$ -- see
Fig.~\ref{marais_nrj}(a). We note that $\tilde{X}_{max}(t)$ is time-dependent
and should not be confused with $X_{max}$ shown in
Fig~\ref{marais_max}(c). The quantities are related via:
$X_{max}=\max_t\tilde{X}_{max}(t)$.
We find that the ratio of $E_{tot}/E_{half}$ remains approximately constant
($\approx 2$) in time, which means that the measurement window is sufficiently
large to capture the dynamics of the perturbation growth and decay before the
effect of the flow structures advected away from the downstream boundary of
the measurement window becomes significant.

The time evolution of the energy is shown in Fig.~\ref{marais_nrj}(b). The
value of $E_{tot}$, as well as $E_{half}$ and $E_{tot}-E_{half}$, are
shown. All energies are all normalized by $E_0$, the value of $E_{tot}$ at the
first measured instance following the impulse.
The perturbation energy initially undergoes growth until $t \approx
50T_{adv}$, at which point the energy decays. This is precisely the transient
growth dynamics expected of convective instabilities in inhomogeneous media.

\begin{figure*}
\centering
 \includegraphics[width=0.48\linewidth]{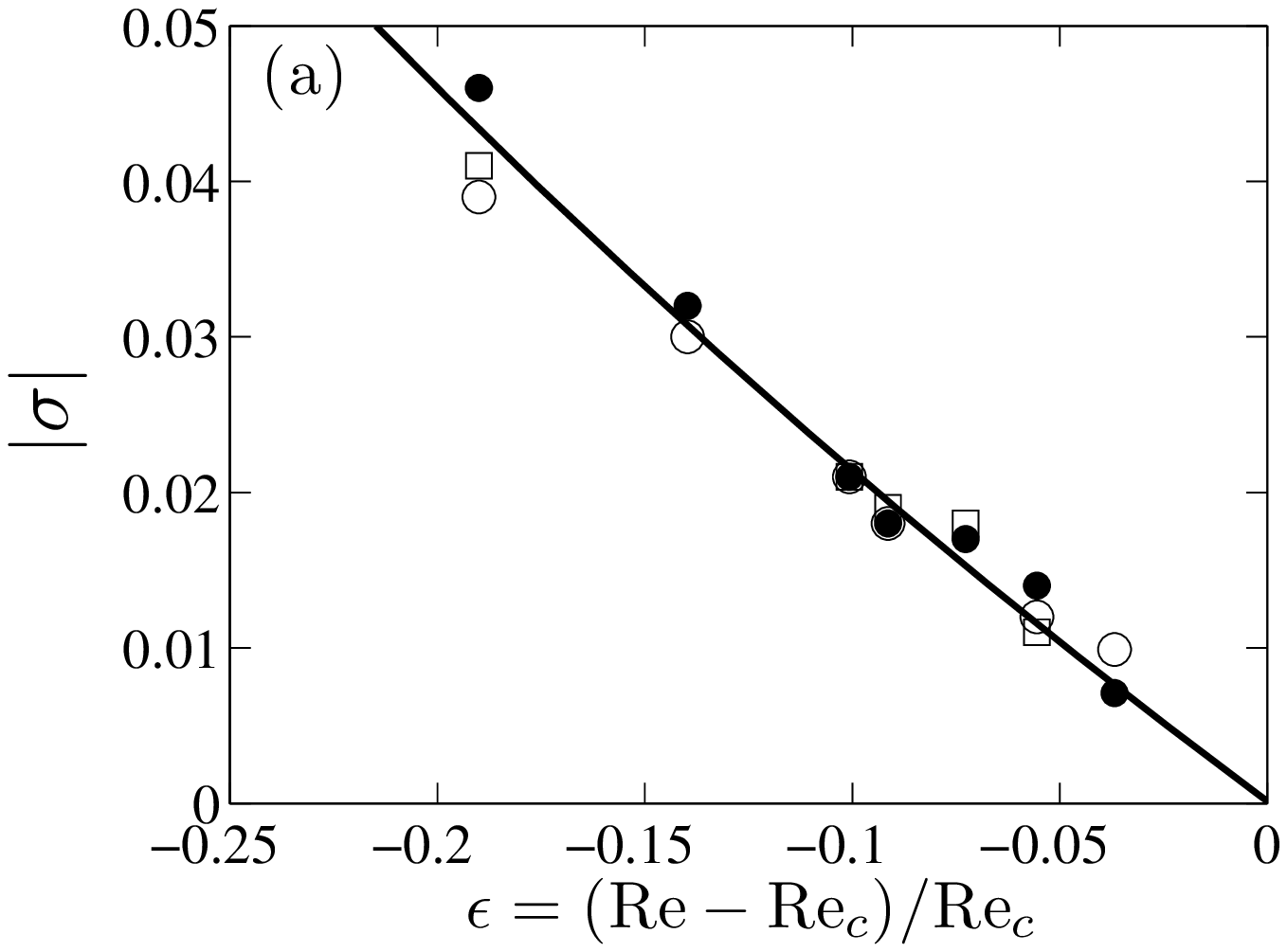}
 \includegraphics[width=0.48\linewidth]{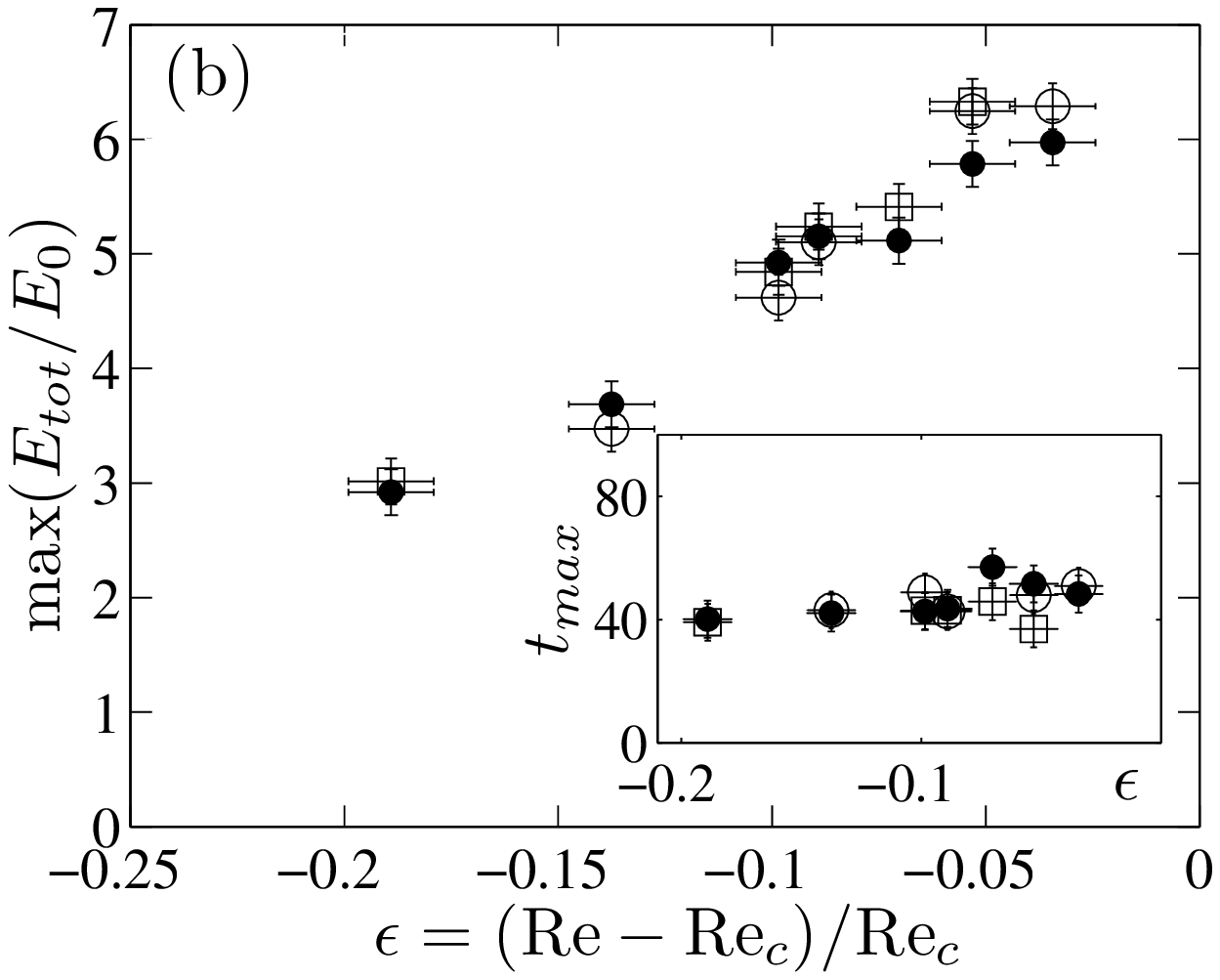}
  \caption{(a) Asymptotic decay rate of perturbations [as seen in
   Fig.~\ref{marais_nrj}(c)] as a function of reduced Reynolds number
   $\epsilon$.  Experimental points are labeled as $\bullet$: low
   perturbation, $\circ$: medium perturbation, $\square$: strong
   perturbation. The curve shows decay rates from linear stability
   computations. (b) Maximum transient energy
   gain and (inset) time at which this maximum occurs as, a function of the
   reduced Reynolds number $\epsilon$. Experimental points are labeled as
   $\bullet$: low perturbation, $\circ$: medium perturbation, $\square$:
   strong perturbation.}
  \label{marais_tg_vs_eps}
\end{figure*}

The late time behavior of the energy corresponds to the exponential decay of
the least stable normal mode in a stable
region,~e.g.~\cite{HernandezPacheco2002,GiannettiLuchini:2007}. As the system
approaches the absolute instability threshold ($\epsilon\rightarrow 0$) this
decay becomes slower. This can be be seen in Fig.~\ref{marais_nrj}(c), where
the decay rate is given by the slope of the curves of $\log E_{tot}$ and more
quantitatively in Fig.~\ref{marais_tg_vs_eps}(a) where the asymptotic decay
rate is plotted as a function of reduced Reynolds number.
For comparison, decay rates from linear stability
computations~\cite{Barkley:2006} shown with a solid curve.  The agreement is
excellent.

The energy growth at short times can be examined using the maximum of the
energy $E_{max}$ and the time for which this maximum is reached $t_{max}$ as a
function of $\epsilon$. See Fig.~\ref{marais_tg_vs_eps}(b).  Consistently with
the approach of the absolute instability threshold, $E_{max}$ increases when
$|\epsilon|\rightarrow0$.  The time $t_{max}$ where the maximum energy is
reached remains approximately constant.

\section{Discussion and conclusions}

The convectively unstable wave packets produced by an impulsive perturbation
in a subcritical cylinder wake have been studied experimentally. Velocity
field measurements obtained by PIV have permitted us to characterize
quantitatively the instability wave, shedding light on points that
remained not clearly analyzed in the literature. Firstly, probing the
perturbation of the velocity field due to the instability shows unambiguously
that there is a well-defined maximum of the perturbation amplitude in the
wake. This differs significantly from the picture given by the streakline
visualizations usually used to illustrate the convective instability
\cite{LeGal2000}, which distort the observation of the actual growth and
decay.  The value of this maximum and its position downstream the cylinder
depend on the Reynolds number and, less markedly, on the strength of the
perturbation, whereas its position in time remains constant.  In addition, the
evolution of the wave packet has been characterized with respect to the
Reynolds number using the leading and trailing fronts as well as a typical
group velocity. The measured velocities are consistent with
the transition from a convective to an absolute global instability as the
Reynolds number increases towards the B\'enard-von K\'arm\'an instability
threshold (see Fig.~\ref{marais_spatio_velocities_vs_epsilon}).

The analysis of the transient energy growth associated to the instability also
deserves a final comment.  The qualitative features of the temporal evolution
of the energy agree with the standard picture of transient growth due to
convective instability in inhomogeneous media, i.e. a short-time algebraic
growth followed by an exponential decay at late times. However, the measured
values of $\max(E_{tot}/E_0)$ are remarkably low (always less than 10) when
compared to the values obtained by numerical computations (of order $10^3$,
see e.g. \cite{Abdessemed2009,Cantwell2010b}).  The main reasons for this
discrepancy are most likely the fact that in the experiment one does impose
an \emph{optimal} perturbation and also the fact that extracting $E_0$, the
initial energy of the perturbation, is experimentally quite difficult.
This discrepancy raises the question about the pertinence of the
energy gain (ubiquitous parameter in transient growth studies) as the most
appropriate quantity to use for comparison between theory and experiments.

\begin{acknowledgments}

DB gratefully acknowledges support from the Leverhulme Trust and the Royal
Society.

\end{acknowledgments}

\newcommand{\noopsort}[1]{} \newcommand{\printfirst}[2]{#1}
  \newcommand{\singleletter}[1]{#1} \newcommand{\switchargs}[2]{#2#1}

\bibliographystyle{pf}

\end{document}